\documentclass[aps,prb,twocolumn,showpacs,superscriptaddress]{revtex4-2}  
\usepackage{graphicx}  
\usepackage{bm}        
\usepackage{amssymb}   
\usepackage{amsmath}   
\usepackage[colorlinks=true,allcolors=blue,breaklinks]{hyperref}
\usepackage{upgreek}
\usepackage{chemmacros}[2014/01/24]
\usepackage{color}
\usepackage{booktabs,multirow}

\newcommand{\be}{\begin{equation}}

\newcommand{\ee}{\end{equation}}
\newcommand{\bea}{\begin{eqnarray}}
\newcommand{\eea}{\end{eqnarray}}

\renewcommand{\vec}[1]{{\bf #1}}

\renewcommand{\tilde}{\widetilde}

\graphicspath{{figures/}}

\def\afflux{Department of Physics and Material Science, University of Luxembourg, L-1511 Luxembourg, Luxembourg}
\def\affbrin{Research Center for Quantum Physics, National Research and Innovation Agency, South Tangerang, Indonesia}
\def\affugm{Department of Physics, Universitas Gadjah Mada, Sekip Utara BLS 21 Yogyakarta 55281, Indonesia.}
\def\affui{Department of Physics, Faculty of Mathematics and Natural Sciences, University of Indonesia, Depok 16424, Indonesia}
\def\affitba{School of Electrical Engineering and Informatics, Bandung Institute of Technology, Bandung, West Java, Indonesia}
\def\affitbb{Research Center for Nanosciences and Nanotechnology (RCNN), Bandung Institute of Technology, Bandung, West Java, Indonesia}
\def\affits{Department of Physics, Faculty of Science and Data Analytics, Institut Teknologi Sepuluh Nopember, Surabaya 60111, Indonesia.}

\begin{document}
\title{Optimal half-metal band structure for large thermoelectric performance}
\author{Finantius E. M. Rahangiar}
\affiliation{\affugm}
\affiliation{\affbrin}
\author{Adam B. Cahaya}
\affiliation{\affui}
\author{Melania~S. Muntini}
\affiliation{\affits}
\author{Isa Anshori}
\affiliation{\affitba}
\affiliation{\affitbb}
\author{Eddwi H. Hasdeo}
\affiliation{\affbrin}
\affiliation{\afflux}
\date{\today}

\begin{abstract}
    Half-metal ferromagnets were predicted [in IEEE Trans. Mag. 51, 1 (2015)] to give large thermoelectric performance in anti-parallel spin valve configuration.  Despite being metals that suffer from the Wiedemann-Franz law, the additional spin degrees of freedom allow for tuning of the thermoelectric properties due to the spin-valve enhancement factor (SVEF). We test this theory and find a mismatch of parameters that gives large TE performance and large SVEF. As a result, we show that the spin-valve setup is useful only for gapless HMF with initially poor TE performance. To obtain the largest TE performance, one still needs to open the band gap. 
\end{abstract}
O\maketitle
\section{Introduction}
Thermoelectric (TE) materials are capable of transforming heat into electricity. Despite the fact that research on this topic has been conducted for centuries, it is still difficult to obtain TE materials for practical applications. The power factor $PF=S^2\sigma$ and the dimensionless figure of merit $ZT=(PF/\kappa) \, T$ are two parameters used to determine whether a material is suitable for TE purposes. Here, $S$ is the Seebeck coefficient, $\sigma$ is the electrical conductivity, $\kappa$ is the thermal conductivity and $T$ is the average of the hot and cold temperatures.
Much effort has been put into improving both $ZT$ and PF. Thermoelectric performance can be increased by either increasing PF or reducing thermal transport. Common methods for this include using low-dimensional structures ~\cite{PhysRevLett.117, PhysRevB.47.16631, PhysRevB.47.12727, Heremans2013}
and manipulating the band structure~\cite{bandengg} through doping~\cite{MgMnBi} and strain~\cite{acs.langmuir.2c03185,PhysRevApplied.strain}. Recent progress in topological insulators enables a dissipationless edge channel to enhance thermoelectric performance~\cite{hasdeo21-TI, Xu2017}.
The other method by enhancing phonon scattering~\cite{phononscatt} to reduce phonon transport~\cite{phonon, Xie_2024}.

Metals are not ideal TE materials because electrons that carry charge current also carry heat which obeys the Wiedemann-Franz law $\sigma T/\kappa = (3k_B^2)/(\pi^2 e^2)$, where $k_B$ is the Boltzmann constant and $e$ is charge of an electron~\cite{WFL}. In addition, the Seebeck coefficient, which is the ratio of electric field to temperature gradient ($S=-\nabla V/\nabla T$), is very small in metals, resulting in a $ZT$ value that is much less than one. Despite the small TE efficiency of metals, the power factor can be large. Currently, Mg$_3$Bi$_2$-based materials possess one of the largest PF which are about $20$ $\mu\rm W/cmK^2$~\cite{mao2019high}. This large PF is useful when there is an unlimited heat source and output power is prioritized over efficiency. 

\begin{figure}[b]
\centering
\includegraphics[width=8cm]{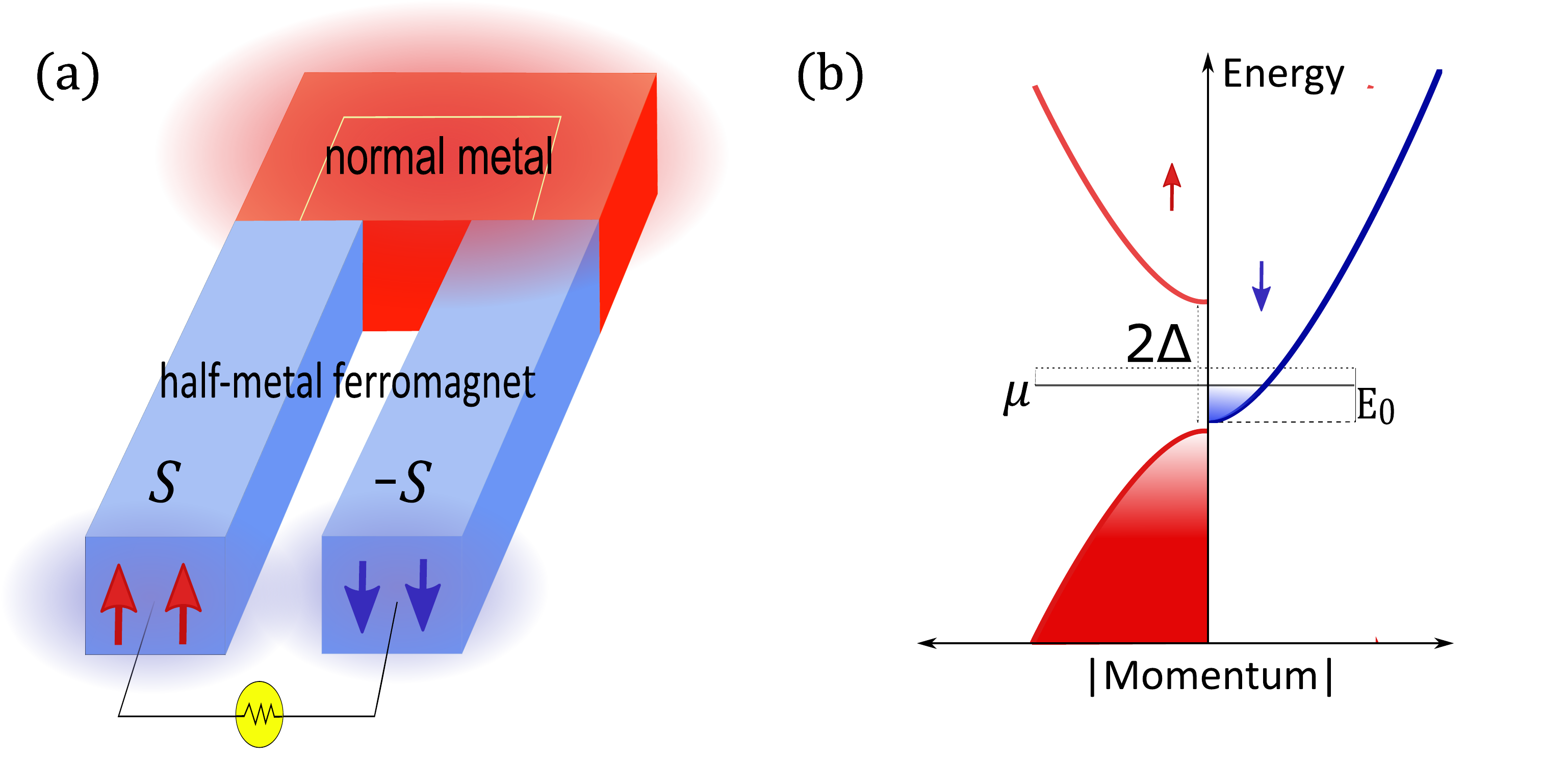}
\caption{(a) Spin valve configuration of half-metal ferromagnetic, and (b) band structure of half-metal ferromagnetic using parabolic band model}
\label{fig1}
\end{figure}

When spin degrees of freedom are involved in TE transport, both the PF and $ZT$ of metals can be further increased~\cite{adam15}. In Fig.~\ref{fig1}(a), half-metal ferromagnets with opposite spin majority in two legs are connected by a normal metal. The Seebeck coefficients in both legs are assumed to be of equal and opposite sign ($S$ and $-S$). In this antiparallel setup, the spin current cannot traverse from one leg to the other, creating additional resistance that increases the Seebeck coefficient \cite{PhysRevB.79.174426} and consequently $ZT$ and PF~\cite{adam15,CrX}. This effect is referred to as the spin-valve enhancement factor (SVEF).

In the previous work, we considered half metallic two-dimensional (2D) chromium pnictides~\cite{CrX}. We predict the enhancement of $ZT$ for CrAs, CrSb, and CrBi but not in CrP.  In some values of the Fermi energy, the SVEF is less than unity, indicating that no enhancement has been achieved. Therefore, it is necessary to find the optimal band structure and parameters to achieve the largest PF and $ZT$ in this system. In this work, we start with the simplest model of the half-metal band structure and find the optimal parameters to obtain the largest SVEF, PF, and $ZT$. We describe a half-metal as spin-polarized bands comprising a single metallic band (blue line in Fig.~\ref{fig1}(b))  and two insulating bands (red lines in Fig.~\ref{fig1}(b)) with a band gap $2\Delta$  and opposite spin orientation. We model the metallic band as a single parabolic band that has a finite density at equilibrium determined by its band depth $E_0$ measured from the charge neutrality point $\mu=0$. Since we consider 2D materials, the Fermi energy $\mu$ is tunable by a gate voltage. In this work, we search for the optimal parameters $\Delta$, $E_0$, and $\mu$ to obtain the largest SVEF.

\section{Model and Methods}

Here, we model a half-metal ferromagnetic (HMF) system using a parabolic band for the spin-down metal state and two parabolic bands for the spin-up insulating state. The electronic band structures for metal is given by
\bea
E_m(\vec k) &=& \frac{\hbar^2\vec k^2}{2\bar m} -E_0 ,
\eea
while for insulator $E_i$ consists of 
\bea
E_c(\vec k) &=& \frac{\hbar^2\vec k^2}{2\bar m} +\Delta  ,\\
E_v(\vec k) &=& -\frac{\hbar^2\vec k^2}{2\bar m} -\Delta,
\eea
where $E_0$ is the depth of metallic band  and $\Delta$ is half of band gap of insulating bands, as illustrated in Fig.~\ref{fig1}(b). 

We can vary the effective mass $\bar m$ for three bands, however, the TE kernel 
\begin{align}\label{eq:1}
    \mathcal{L}^{(n)}_{j} = \sum_{N_{j}}\int \tau \textbf{v}_{j}^2 (E)\mathcal{D}_{j}(E)(E-\mu)^n \biggl( -\frac{\partial f}{\partial E}\biggl) dE
\end{align}
defined in the linearization of the Boltzmann equation, is independent of mass due to the cancellation of $\bar m$ in ${\bf v}^2$ and $\mathcal D$ [see Appendix A]. Here $j=m,i$ are indices for metal and insulator band respectively with $N_m=m$ and $N_i=c,v$, i.e. number of bands.   $\mathcal{D}_j(E)=\sum_\vec k \delta(E_j(\vec k)-E)$ is the density of states, $\tau$ is the relaxation time
, and $\textbf{v}_{j}=\hbar^{-1}\partial E_{j}/\partial k_x$ is the longitudinal velocity of the electron. TE properties of our system is obtained using the Boltzman transport equation with relaxation time approximation. We argue that a constant relaxation time independent to energy $E$ is a good approximation in the 2D parabolic band due to the constant density of states. Later in Sec.~\ref{SecMismatch}, we also estimate corrections due to an energy-dependent relaxation time $\tau(E)$. 
For simplicity, here we assume negligible phonon contribution to $\kappa$ as shown in previous work~\cite{CrX}.  

Thermoelectric transport coefficients are given by
\begin{align}\label{eq:2}
    \sigma_{j} &= e^2 \mathcal{L}^{(0)}_{j}, \quad j=m,i,\\
\label{eq:3}
   S_{j} &= -\frac{1}{eT}\frac{\mathcal{L}^{(1)}_{j}}{\mathcal{L}^{(0)}_{j}}, \\
\label{eq:4}
    \kappa_{j} &= \frac{1}{T}\biggl( \mathcal{L}^{(2)}_{j}
- \frac{(\mathcal{L}^{(1)}_{j})^2 }{\mathcal{L}^{(0)}_{j}}\biggl),
\end{align}
where $\sigma$, $S$, and $\kappa$ are electrical conductivity, Seebeck coefficient, and electronic thermal conductivity. 
Transport coefficient of the whole system is described by
\begin{align}
\label{eq:5}
    \sigma_{t} &= \sigma_{i} + \sigma_{m},\\
\label{eq:6}
     S_{t} &= \frac{\sigma_{i}S_{i}+\sigma_{m}S_{m}}{\sigma_{i}+\sigma_{m}},\\
\label{eq:7}
    \kappa_{t} &= \kappa_{i} + \kappa_{m}.
\end{align}

Using transport coefficient of the system, we can define the figure of merit and power factor as well as their corresponding spin-valve enhanced values as follows:
\begin{align}\label{eq:8}
    ZT_{v} &=  \chi ZT= \chi \frac{\sigma_{t} S_{t}^2}{\kappa_{t}},\\
\label{eq:9}
    PF_{v} &=  \chi PF= \chi \sigma_{t} S_{t}^2,
\end{align}
where $\chi$ is the spin-valve enhancement factor (SVEF) due to non-parallel configuration shown in Fig.~\ref{fig1}(a)~\cite{adam15}. SVEF is given by:
\be
 \chi = \frac{(1-PP')^2}{1-P^2},
\ee
where $P$ and $P'$ are respectively related to spin polarization of charge and heat, 
\begin{align}\label{eq:10}
    P&=\frac{\sigma_{m}-\sigma_{i}}{\sigma_{m}+\sigma_{i}},\\
\label{eq:11}
    P'&=\frac{\sigma_{m}S_{m}-\sigma_{i}S_{i}}{\sigma_{m}S_{m}+\sigma_{i}S_{i}}.
\end{align}
Parenthetically, we neglect the spin-orbit coupling so that the polarizations can be given simply by Eqs.~\eqref{eq:10} and~\eqref{eq:11}.  In the presence of spin orbit coupling, Eqs.~\eqref{eq:5}--~\eqref{eq:7} remain intact while for Eqs.~\eqref{eq:10} and~\eqref{eq:11} should be defined through the spin projection procedure.

\section{Results}

\subsection{Thermoelectric Coefficients}

\begin{figure*}[t]\centering
\includegraphics[width=12cm]{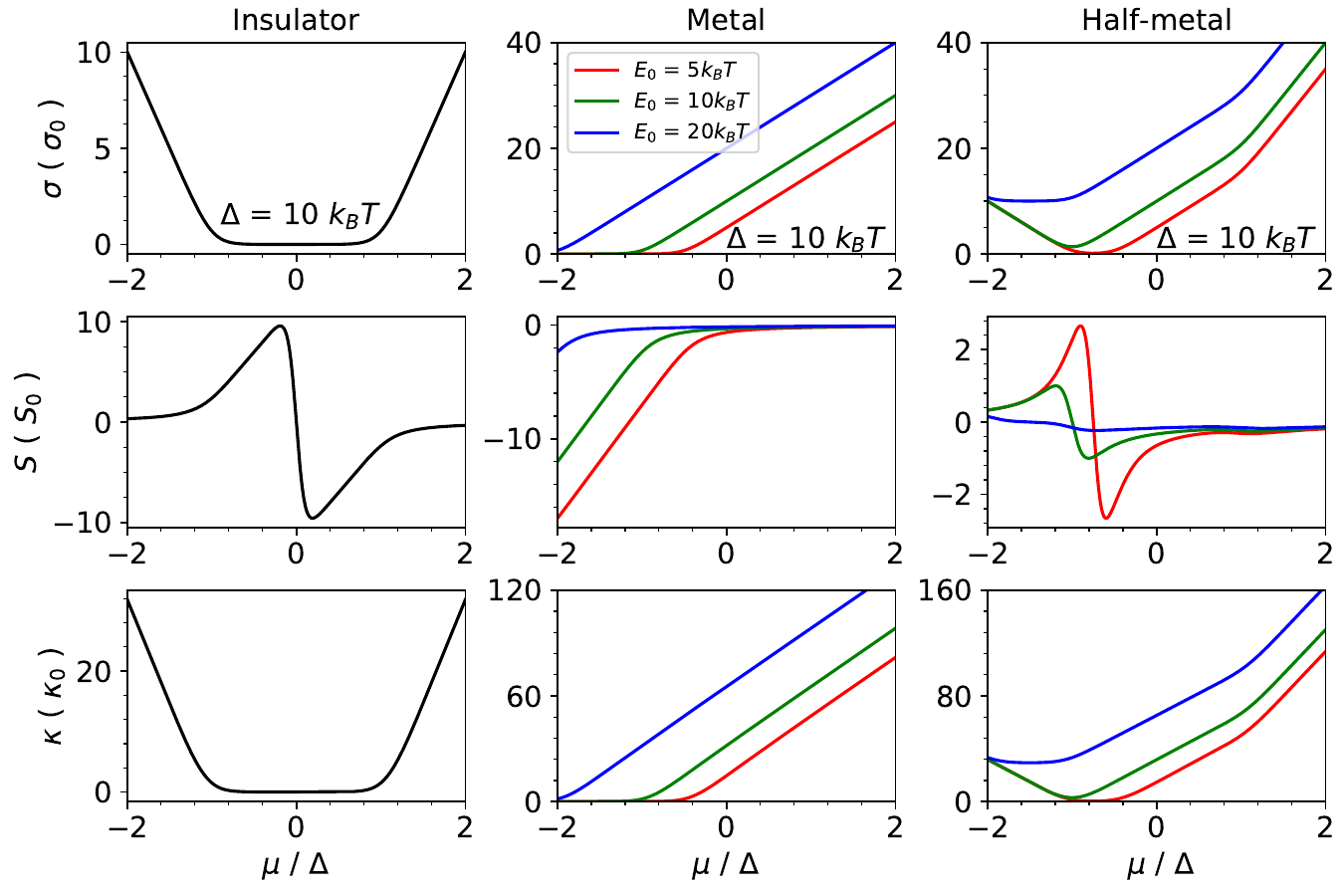}\caption{Thermoelectric coefficients of insulator, metal, and half metal. Electrical conductivity, Seebeck coefficient, and thermal conductivity as a function of $\Delta$ and varying $E_0$.}\label{fig2}
\end{figure*}

In figure \ref{fig2}, we show separately the thermoelectric coefficients ($\sigma$, $S$, and $\kappa$) of the spin-down state (insulator), spin-up state (metal) and the total contribution of both spins as a function of the Fermi energy $\mu$. We fix the band gap and vary the depth of the metallic band $E_0$.  
TE coefficients are plotted in units of $\sigma_0= \tau_0 k_BTe^2/\pi L\hbar^2$, $S_0=k_B/e$, and $\kappa_0=\tau_0 k_B^3 T^2/\pi L \hbar^2$ 
where $\tau_0$ is the relaxation time and $L$ is confinement length. Using a typical confinement length $L=1\ {\rm nm}$, $T=300\ \rm K$ and $\tau_0=1\ {\rm fs}$, we obtain $\sigma_0 = 3043.1$  ${\rm S/m}$, $S_0=86.1\ \rm{\mu}\rm{V}/\rm{K}$, and  $\kappa_0= 677.7 $ $\rm{\mu W}/\rm{cm K}$.

As summations of contributions from the insulating and metallic bands, the electric $\sigma_t$ and thermal $\kappa_t$ conductivities of the half-metallic band possess values higher than those of each individual spin. Meanwhile, the Seebeck coefficient of the half-metallic band shows an insulating-- or metallic--like character depending on the value of $E_0$ vs. $\Delta$.  For $E_0$ is larger than $\Delta$, the Seebeck coefficient of half-metal is monotonic and therefore has a similar character to metal. On the other hand, for $E_0$ is smaller than $\Delta$, the Seebeck coefficient shows an insulating-like character with the peak positions being shifted from the charge neutrality point $\mu=0$. However, the combined Seebeck coefficient values on the HMF are lower than those of individual bands given by Eq.~\eqref{eq:6}.


\subsection{$ZT$ and PF of half metals}

\begin{figure}[b]
\centering
\includegraphics[width=9.0cm]{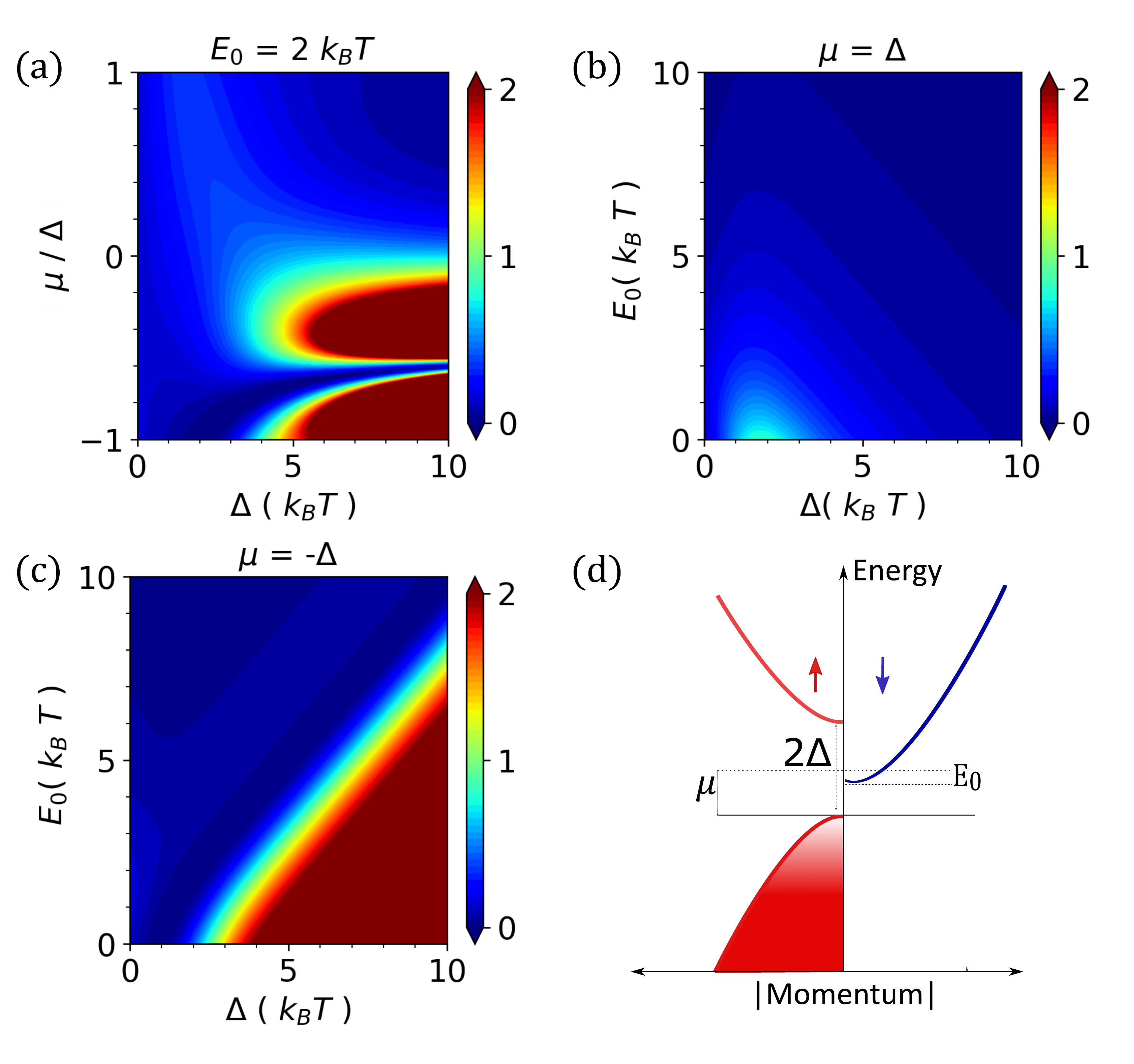}
\caption{
(a) The 2D plot of ZT as functions of Fermi energy $\mu$ and $\Delta$ for $E_0=2\ k_BT$. (b) and (c) are $ZT$ for $\mu = \Delta$ and $\mu = -\Delta$, respectively, plotted as a function of $E_0$ and $\Delta$. (d) Optimal band structure to achieve optimal $ZT$.}
\label{fig5}
\end{figure}

Based on the results in Fig.~\ref{fig2}, one can calculate the dimensionless $ZT$ and PF using Eqs.~\eqref{eq:5}--~\eqref{eq:7}. In Fig.~\ref{fig5}(a) we show the 2D plot of $ZT$ as functions of $\mu$ and $\Delta$ for $E_0=2\ k_B T$. In the negative $\mu$, we obtain a very large $ZT$ above $2$ (extended color bar). Taking a vertical cut of $ZT$ along $\Delta=6\ k_BT$, one observes double peak structures of ZT as a function of $\mu$. These double peaks originate from $S_t^2$. Focusing on $\mu=\pm\Delta$ [Fig.~\ref{fig5} (b,c)], we scan $ZT$ over the $\Delta-E_0$ space. At $\mu=\Delta$, $ZT$ is generally less than $1$. On the other hand, at $\mu=-\Delta$, $ZT$ can be higher than $2$ as long as $\Delta > E_0+\xi$, with $\xi\approx 3\ k_BT$ [see Fig.~\ref{fig5}(c)]. This means that the ideal band structure to achieve the highest $ZT$ is slightly gapped out with a band gap of approximately $\xi$ [see Fig.~\ref{fig5}(d)]. We note that the region with large $ZT$ extends to large values of $E_0$ and $\Delta$. However, for large $\Delta$ one needs large hole doping to reach $\mu=-\Delta$.

\begin{figure}[t]
\centering
\includegraphics[width=8.5cm]{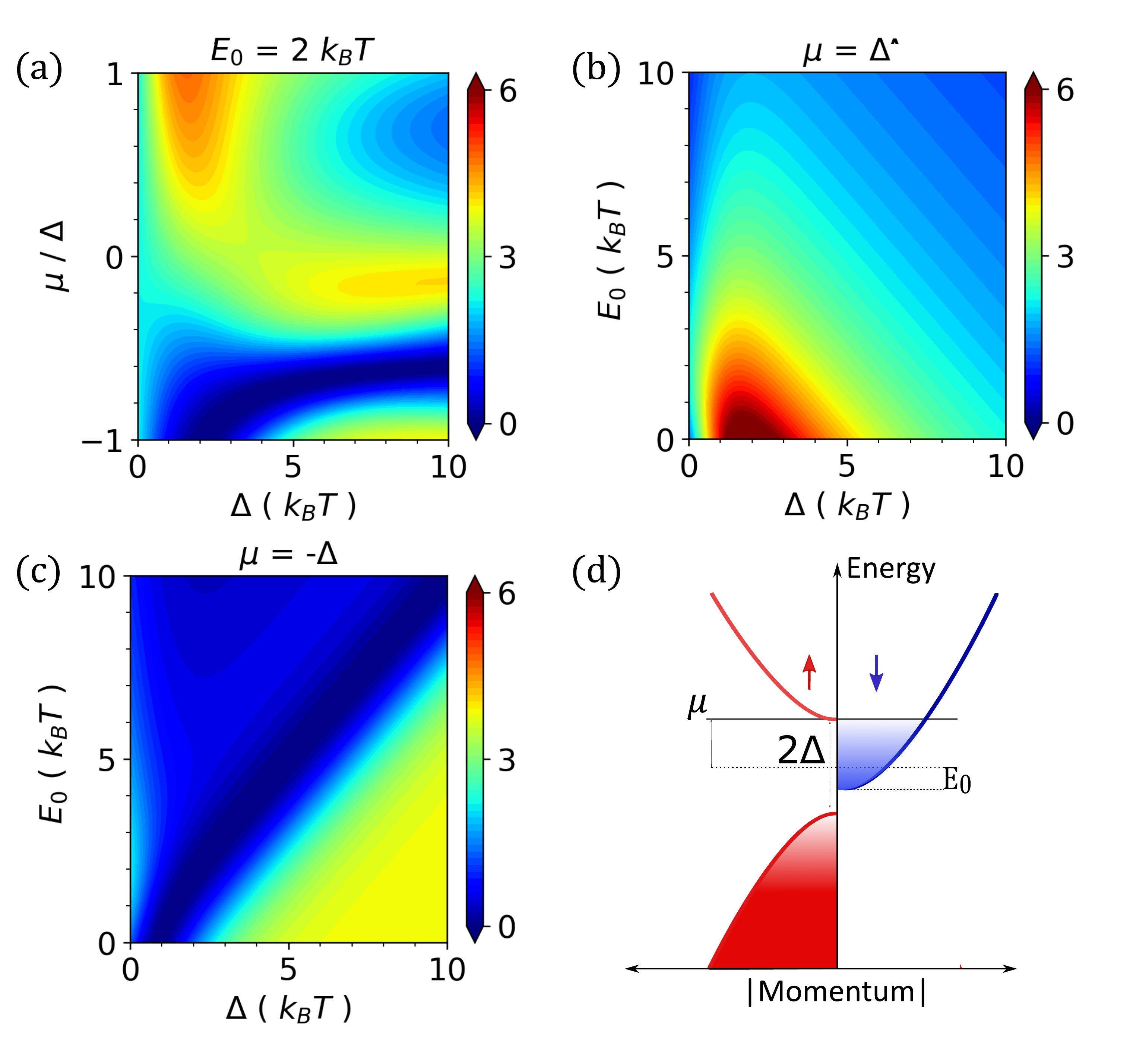}
\caption{
(a) The 2D plot of PF as functions of $\mu$ and $\Delta$ for $E_0=2\ k_BT$. (b) and (c) are $PF$ for $\mu=\Delta$ and $\mu=-\Delta$, respectively, plotted as a function of $E_0$ and $\Delta$. (d) Optimal band structure to achieve the largest PF. }
\label{fig6}
\end{figure}

Next, we turn our attention to PF. The HMF power factor reaches a very high value at $\mu=\Delta$ and for very low $\Delta$ [see Fig.~\ref{fig6}(a)]. Focusing on $\mu=\Delta$, we can obtain optimal $\Delta$ and $E_0$ in Fig.~\ref{fig6}(b). Small $\Delta\approx 2-3\ k_BT$ and small $E_0<3\ k_BT$ produce very large $\textrm{PF}\approx 6 \textrm{PF}_0$ where $\textrm{PF}_0=0.23 \times \bar \tau\ \mu{\rm W/cm K^2}$, with $\bar\tau$ is the relaxation time in fs. To reach the current record of PF, one requires a relaxation time of about $10$ fs, which is a moderately clean sample. 
For $\mu=-\Delta$, one can obtain a moderately large PF in the region where $ZT$ is large [see Fig.~\ref{fig6}(c)]. The optimal band structure for the largest PF is summarized in Fig.~\ref{fig6}(d). In this case, $\mu=\Delta$ and very small $E_0$ and $\Delta$, which means that we need to open the HMF band gap about $\Delta$.

\subsection{Spin Valve Enhancement Factor (SVEF)}

\begin{figure}[t]
\centering\includegraphics[width=8.5cm]{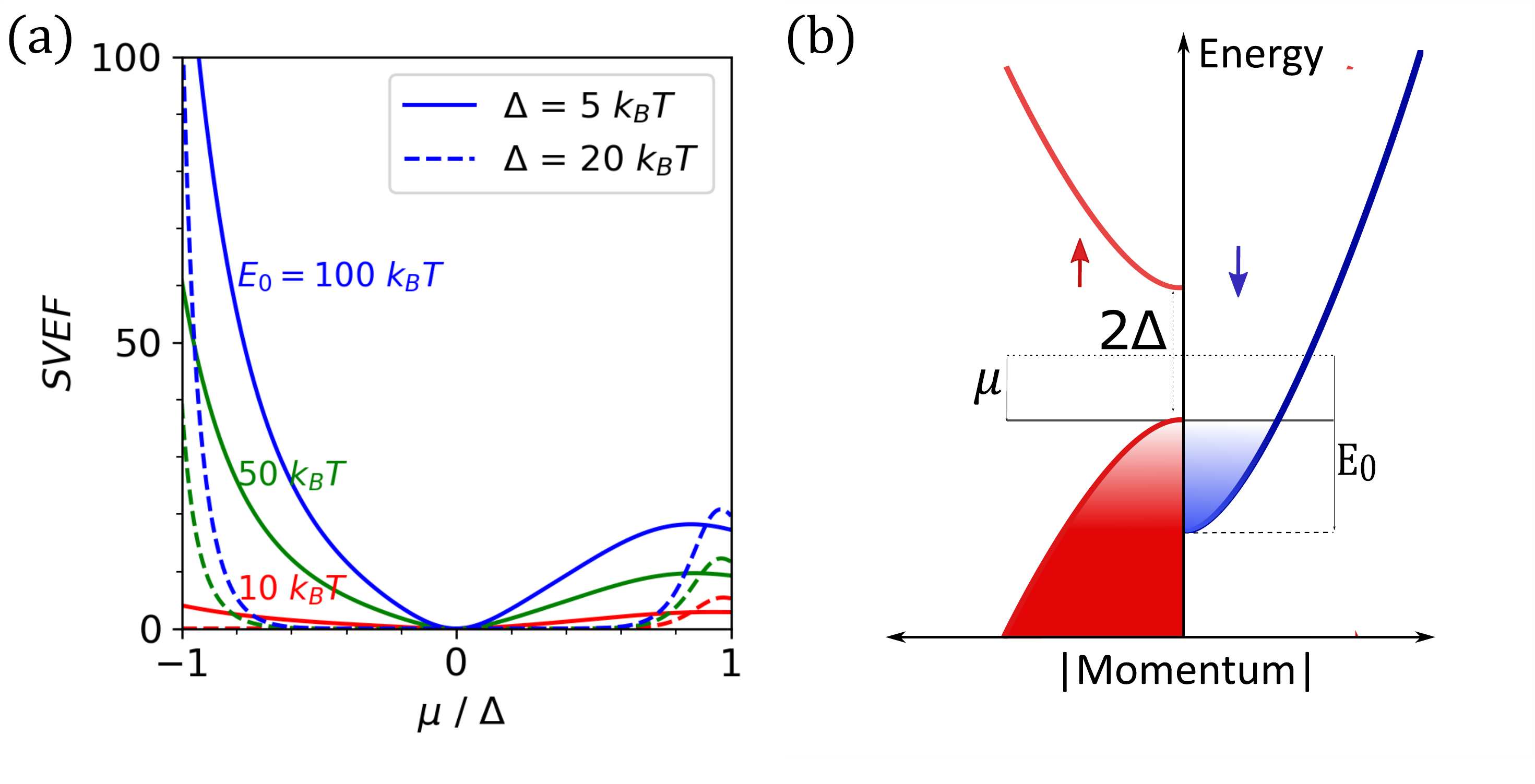}
\caption{(a) Spin valve enhancement factor as function of Fermi energy ($\mu/\Delta$) for various values of $E_0$. The dashed lines are for $\Delta=20\rm{k_BT}$ and the solid lines are for $\Delta=5\rm{k_BT}$.(b) Optimal band structure for large SVEF}\label{fig3}
\end{figure}

In Eqs. (\ref{eq:8}) and (\ref{eq:9}), thermoelectric performance is enhanced when $\textrm{SVEF} > 1$. Therefore, SVEF is expected to improve the TE performance of half-metals.  Figure \ref{fig3}(a) shows the SVEF of HMF as a function of the Fermi energy $\mu$ from -$\Delta$ to $\Delta$. The solid lines are for  $\Delta = 5\ k_B T$ and the dashed lines are for $\Delta=20\ k_B T$, with each color represent different metallicity.  The results show that a large band gap yields  small values of SVEF at low Fermi energy and drastically increases near the band edge. SVEF increases greatly with the increase of $E_0$. In the limit of $E_0\gg \Delta$, charge polarization is dominated by the metal sector which gives $P\approx 1$ [see Eq.~\eqref{eq:10}] in the entire range $\mu$  between $[-\Delta,\Delta]$. On the other hand, the values of $P'$ can deviate from unity due to the large contribution of $S_i$ near the band edges. The degrees of polarization $P$ and $P'$ are proportional to $E_0/\Delta$. For small $\Delta$, thermal carriers are excited above the gap  giving the smeared profile of SVEF inside the gap (solid lines). The bigger $E_0$, the closer $P$ value to unity at the edges, resulting in a linear increase in SVEF.  
The highest SVEF is achieved at $\mu = -\Delta$ because in these parameters $S_i$ and $S_m$ give opposite signs [see Eq.~\eqref{eq:11}] with $P'=3$ and $P\approx 1$ for $E_0\gg \Delta$. For $E_0=100\ k_BT$, SVEF reaches 100 at $\mu=-\Delta$ and about 20 at $\mu=\Delta$.
Fig.~\ref{fig3}(b) illustrates the optimal band structure for the largest SVEF.

\begin{figure}[t]
\centering\includegraphics[width=8.5 cm,height=8.5cm]{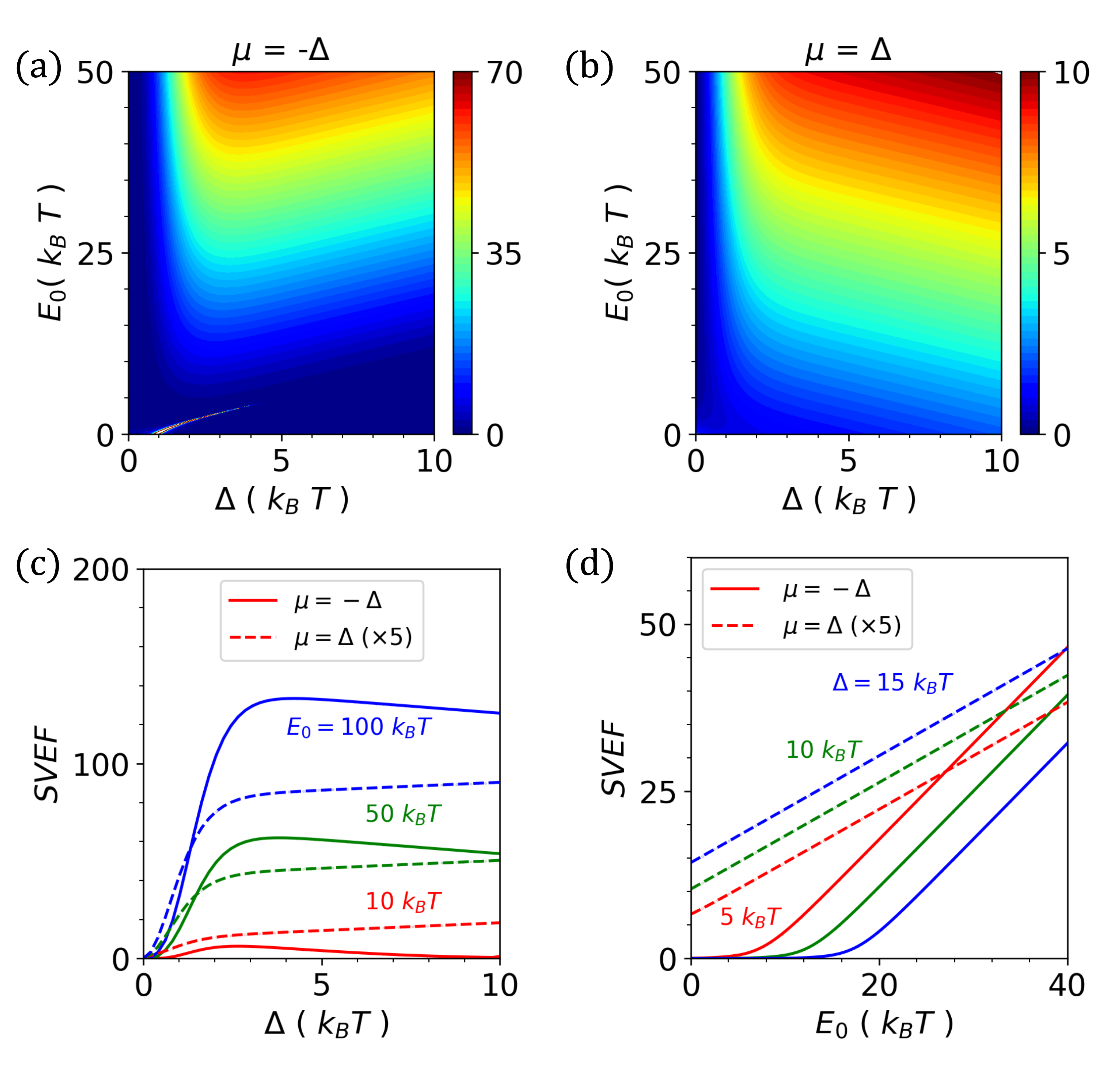}\caption{SVEF of HMF in contour plots at two different Fermi energies (a) $\mu=-\Delta$ and (b) $\mu=\Delta$ as functions of $E_0$ and $\Delta$. (c) SVEF as a function of $\Delta$ for several values of $E_0$ and (d) SVEF as a function of $E_0$ for several values of $\Delta$. In (c) and (d), the solid lines are for  $\mu = -\Delta$ and the dashed lines are for $\mu=\Delta$.( The dashed lines are multiplied by $5$ times so that they can be plotted on the same scale )} 
\label{fig4}
\end{figure}

We focus on two values of $\mu=\pm\Delta$ and find the optimal $E_0$ and $\Delta$ for the best SVEF of the two cases. In Figs.~\ref{fig4}(a) and (b), we show the 2D plot of SVEF as functions of $E_0$ and $\Delta$ for position $\mu$, respectively, at $-\Delta$ and $\Delta$. We take a horizontal cut of Figs.~\ref{fig4}(a) and (b) and plot  SVEF versus $\Delta$ in Fig.~\ref{fig4}(c) for several values of $E_0$. For $\mu=-\Delta$ (solid lines), SVEF shows a peak with an optimal band gap $\Delta$ between $2-3$ $k_B T$. Meanwhile, for $\mu=\Delta$ (dashed lines), the SVEF increases monotonically as a function of $\Delta$. Taking a vertical cut of Figs.~\ref{fig4}(a) and (b), we show that SVEF increases monotonically as a function of $E_0$ at both positions of $\mu$ [see Fig.~\ref{fig4}(d)].  However, at $\mu=\Delta$ ($\mu=-\Delta$), the SVEF is proportional (inversely proportional) to $\Delta$.



As presented in Figs.~\ref{fig5} and \ref{fig6}, high TE performances  require a very small value of $E_0$. On the other hand, as shown in Figs.~\ref{fig3} and~\ref{fig4}, SVEF prefers large $E_0$. These different conditions indicate the mismatch of parameters to obtain maximum SVEF with the parameters that contribute to large $ZT$ and PF. For that reason, we cannot get both of them at the same time.   2D ferromagnetic insulators, for example $\rm CrI_3$, which have a large band gap around $1.89$ $\rm{eV}$ ($ZT=1.57$ at $T=900$) $\rm{K}$~\cite{Sheng_2020}  
 thus do not have large SVEF and PF. $ZT$ can be potentially large, but phonon thermal conductivity might hamper it. On the other hand, cromium pnictide such as $\rm{CrBi}$ is an HMF and has large $\Delta$ and $E_0$. This material has a large SVEF but the initial $ZT$ is small~\cite{CrX}.

\section{TE Performance of spin-valve thermocople}

We have shown that spin valve setup in Fig.~\ref{fig1}(a) can, in principle, increase $ZT$ and PF by a factor of hundreds as shown in Figs.~\ref{fig3} and~\ref{fig4}. In Sec.~\ref{SecZTvPFv}, we discuss which combination of $E_0$, $\Delta$, and $\mu$ give the largest $ZT_{\rm v}$ and PF$_v$. However, the parameters that give large $ZT$ and PF do not match those of large SVEF (cf. Fig.~\ref{fig4}(a) with Figs.~\ref{fig5}(c) and \ref{fig6}(c)). The mismatch is further elaborated in Sec.~\ref{SecMismatch}.

\subsection{$ZT$ and PF with spin valve enhancement factor}
\label{SecZTvPFv}
\begin{figure}[t]
\centering\includegraphics[width=8.5cm]{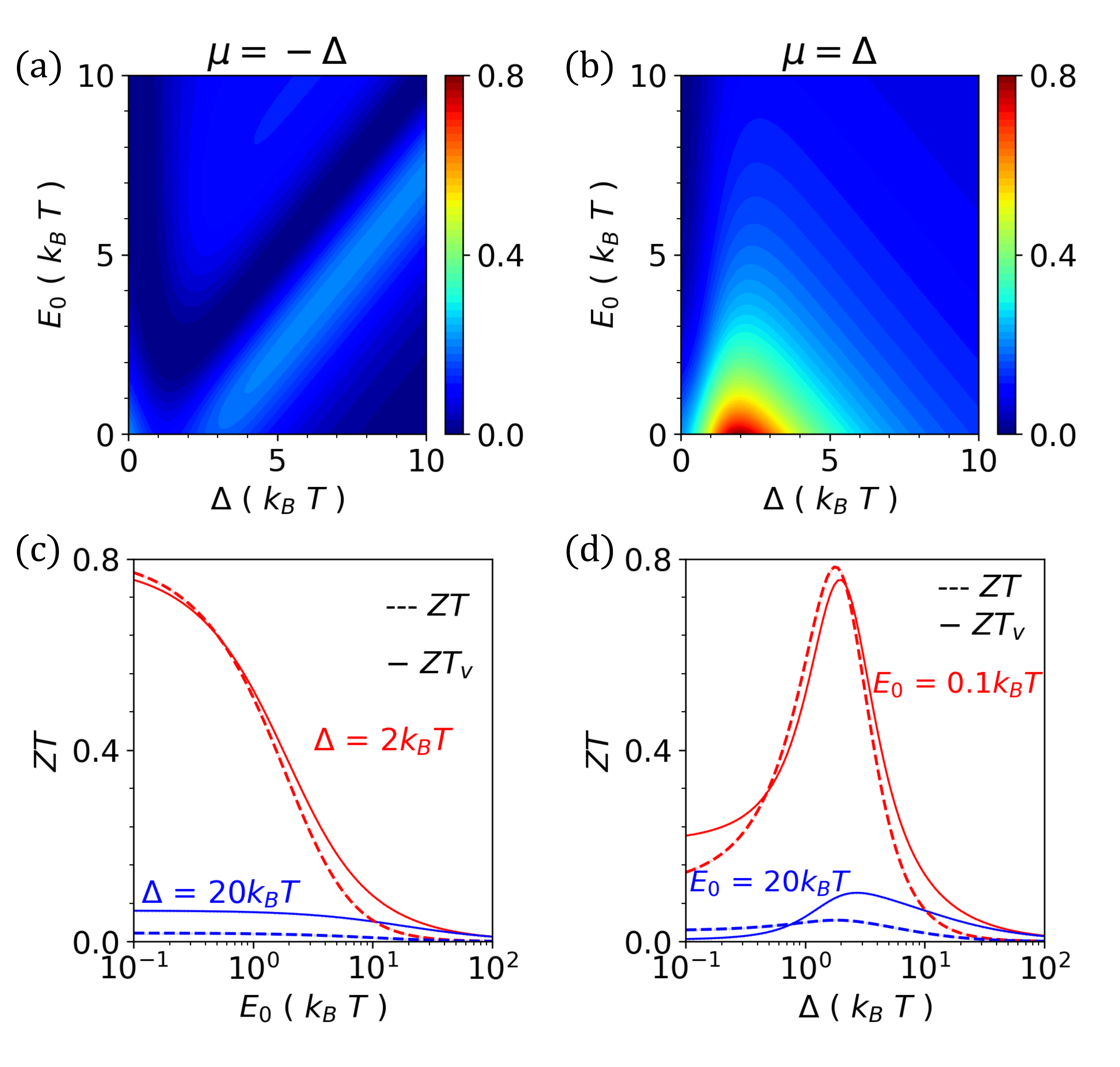}\caption{
(a) and (b) are 2D plot of $ZT$ on spin valve at $\mu = -\Delta$ and $\mu = \Delta$, respectively, as functions of $\Delta$ and $E_0$. (c) and (d) show $ZT$ as functions of $E_0$ and $\Delta$, respectively, at $\mu = \Delta$.}\label{fig7}
\end{figure}

We examine the values of $ZT_{\rm{v}}$ at two different Fermi energies, $\mu=-\Delta$ and $\mu=\Delta$, shown in Figs.~\ref{fig7}(a) and (b). It should be noted that optimal $ZT_{\rm{v}}$ is achieved at $\mu = \Delta$, despite the optimal SVEF occurring at a negative Fermi energy. The maximum value of $ZT_{\rm v}$ is approximately 0.8, which is obtained when $\Delta$ is equal to $2\rm{k_BT}$ and small $E_0$.

In Fig.~\ref{fig7}(c), we plot $ZT_{\rm{v}}$ (solid lines) and compare it with the initial $ZT$ (dashed lines) as a function of log-scaled $E_0$ for $\Delta=2\ k_BT$ (red lines) and $\Delta=20\ k_BT$ (blue lines). The enhancement by spin valve setup, indicated by the difference between the solid and dashed line is small at small band gap and increases a little bit as $E_0$ increases. On the other hand, for larger $\Delta$,  $ZT_{\rm{v}}$ enhances by a factor of $\sim 4$ although the overall value of $ZT_{\rm v}\approx 0.06 $ is small. 

Figure \ref{fig7}(d) shows $ZT$ as a function of log-scaled $\Delta$, which indicates that the optimal $ZT_{\rm{v}}$ is achieved when $\Delta=2\rm{k_BT}$. At the largest $ZT_{\rm v}$ value,  SVEF is relatively small. The contribution relative to the change of $\Delta$ can be compared with Figure \ref{fig4} (d), where the SVEF for $\mu=-\Delta$ increases as $\Delta$ increases. At lower $\Delta$, SVEF is lower than unity, making $ZT>ZT_{\rm v}$. As $E_0$ increases SVEF also increases but the value of $ZT_{\rm v}$ becomes smaller.

\begin{figure}[t]
\centering\includegraphics[width=8.5cm]{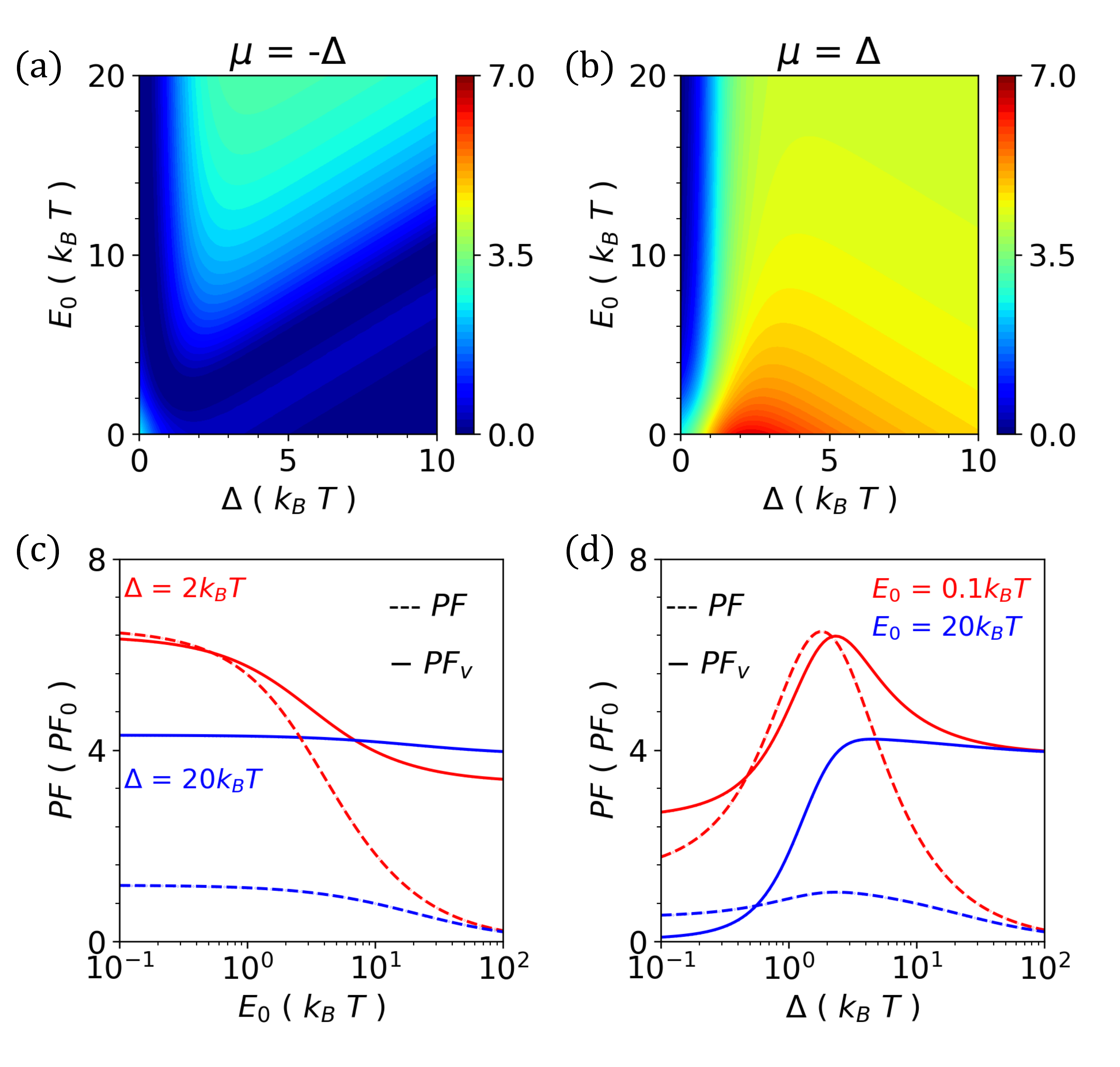}\caption{
(a) and (b) are 2D plot of $\rm{PF}$ on spin valve at $\mu = -\Delta$ and $\mu = \Delta$, respectively, as functions of $\Delta$ and $E_0$. (c) and (d) show the power factor as functions of $E_0$ and $\Delta$, respectively, at $\mu = \Delta$.}\label{fig8}
\end{figure}

Similarly to $ZT$, we also calculate the power factor of the spin valve configuration ($\rm{PF}_{\rm{v}}$). Figures~\ref{fig8}(a) and (b) show the $\rm{PF}_{\rm{v}}$ at $\mu=-\Delta$ and $\mu=\Delta$, respectively. Similarly to $ZT_{\rm{v}}$, the optimal $\rm{PF}_{\rm{v}}$ is obtained at $\mu=\Delta$. Despite that, $\rm PF_v$ at $\mu=-\Delta$  is not negligible, thanks to the large SVEF for $E_0\gg\Delta$. In Figs~\ref{fig8}(c) and (d), the SVEF contribution can be seen from the distance between the solid and dashed lines.  In figure \ref{fig8} (c), the same phenomenon as $ZT_{\rm{v}}$ also appears in $\rm{PF}_{\rm{v}}$, where the high $\rm{PF}_{\rm{v}}$ (red lines) has low SVEF. From figure\ref{fig8}(d), high SVEF is found at higher $E_0$, which has low initial power factor ($\rm{PF}$). On the other hand, lower $E_0$ have a higher initial power factor, which is shown by the red line. Figure \ref{fig8}(d) shows that the peak is also found at $\Delta=2\rm{k_BT}$, which indicates the optimal $\Delta$. 

Our model yields an optimal $ZT_{\rm{v}}$ of 0.8 and a power factor $\rm{PF}_{\rm{v}}$ close to 7$PF_0$. The decrease in power factor due to changes in parameters is not as drastic as the decrease in $ZT$.

Based on these calculations, we propose the third scenario, which achieves the optimal figure of merit and power factor. When the Fermi energy is set at the lowest conduction band ($\mu = \Delta$) with a low metallicity ($E_0 \approx 0.1$) and $\Delta$ around 2$k_BT$ (Figure \ref{fig8}), the optimal $\rm{PF}_{\rm{v}}$ and $ZT_{\rm{v}}$ are achieved. These results suggest that a suitable HMF should have a narrow band gap.



\subsection{Mismatch of TE performance and spin-valve enhancement}
\label{SecMismatch}
Figure~\ref{fig3} shows that the large SVEF is achieved at higher HFM metalicity (larger values of $E_0$), meanwhile, to get better TE performance, one needs to open the gap. At first sight, the mismatch between SVEF and TE performance might depend on the fact that we use a constant relaxation time $\tau=\tau_0 (E/k_BT)^r$ with $r=0$. We reveal the universality of our results by taking into account non-zero $r$.  We can simplify the analysis of Eqs.~\eqref{eq:5}--\eqref{eq:11} by writing $\sigma_m=\gamma\sigma_i$ and $\sigma_mS_m=\xi\sigma_i S_i$ with $\gamma>0$ and $\xi$ can be positive or negative to express $P=(1-\gamma)/(1+\gamma)$, $P'=(1-\xi)/(1+\xi)$, $\chi$ (SVEF), and 
$${\rm PF}= \frac{(\sigma_tS_t)^2}{\sigma_t} = \frac{(1+\gamma)^2}{(1+\xi)} \frac{(\sigma_iS_i)^2}{\sigma_i}$$
as functions of $\xi$ and $\gamma$. 
 
\begin{figure}[t]
\centering
\includegraphics[width=1.1\columnwidth]{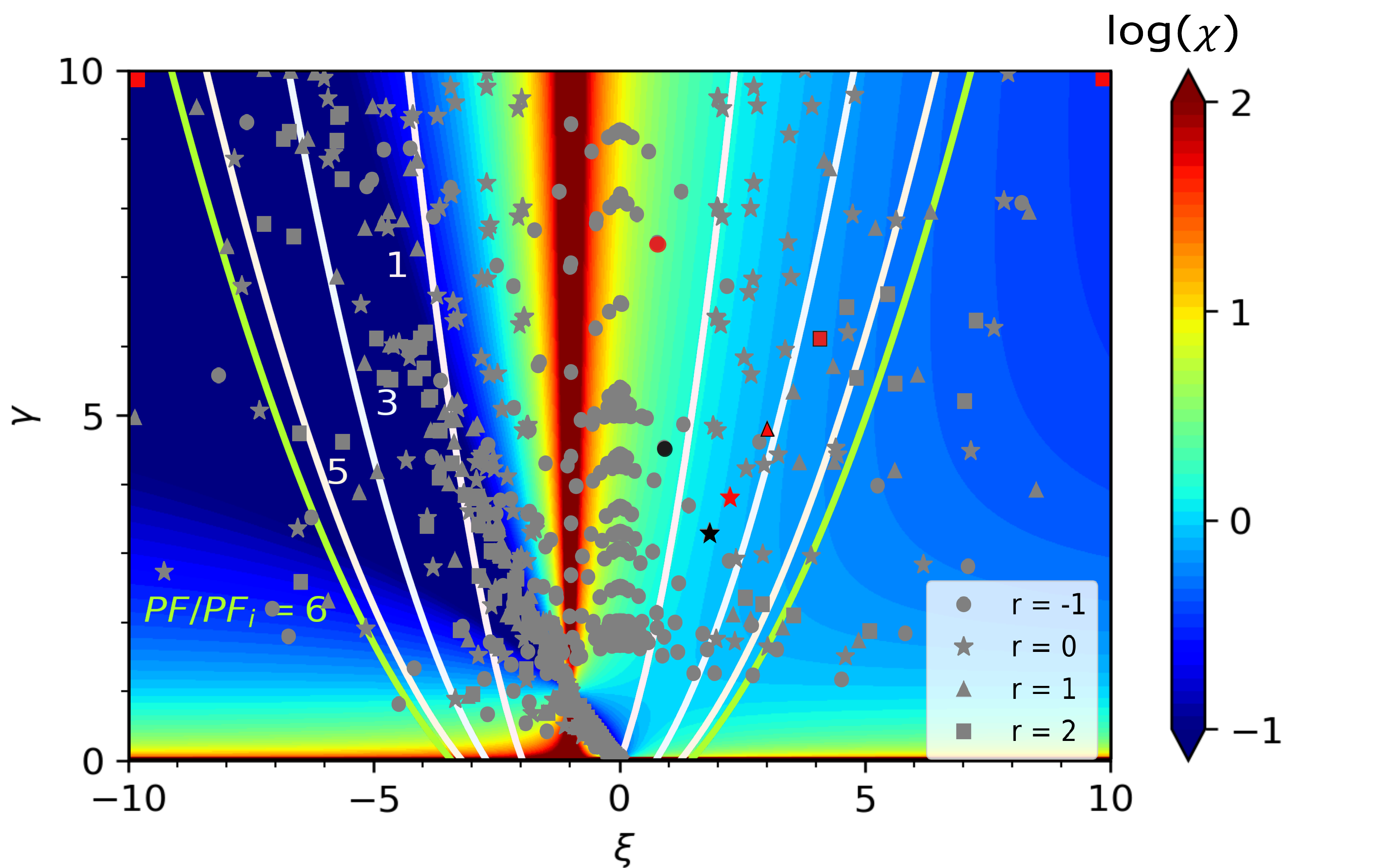}
\caption{
2D plot of $\rm{log}$($\chi$)] as function of $\gamma$ and $\xi$. Line plots represent the region with a certain value of $\rm{PF}/\rm{PF}_i$($=1,3,5,6$) and scatter markers are the data with $r=-1,0,1,2$. Black markers are maximum $\rm{PF}_{v}$ and red markers are the maximum $ZT_v$}
\label{fig10}
\end{figure}

In Fig.~\ref{fig10}, we show $\log(\chi)$ as functions of $\xi$ and $\gamma$ and overlay it with lines that give ${\rm PF}/{\rm PF_i}=1$ up to $6$. From Fig.~\ref{fig10}, we can see that the large SVEF is localized at large $\gamma$ with small $\xi$ or large negative $\xi$ with $\gamma<1$. However, the high ${\rm PF}/{\rm PF_i}$ is found mainly in SVEF $<1$. We also put the spread of data points of $\Delta \in [0,20]\ k_BT$ $\mu\in [-1,1]\Delta$; $E_0 \in [0,20]\ k_BT$; and $r\in [-1,2]$. A typical half metal with large $\gamma$ cannot simultaneously give large PF and SVEF. It is tempting to say that small $\gamma < 1$ and large negative $\xi$ give both large PF and SVEF but it is not true because the lines show the relative value of ${\rm PF}/{\rm PF_i}$ and not the absolute value. The highest values of $\rm PF_v$ and $ZT_{\rm v}$ are shown in black and red, respectively, with different marks indicating different $r$. 
We summarize the optimal $\gamma$ and $\xi$ that gives the maximum $ZT$, PF, $ZT_{\rm v}$, and $\rm PF_v$ in table~\ref{tab:posTE}. 

\begin{table}[h]
    \centering
        \caption{Position $\gamma$ and $\xi$ of maximum TE quantities for different $r$ }
    \label{tab:posTE}
    \begin{tabular}{crrrr}
    \hline
    \hline
\multirow{2}{*}{$r$} & \multicolumn{4}{c}{$(\gamma,\xi)$ } \\
\cline{2-5}
 &  max $ZT$ & max PF & max $ZT_{\rm v}$ & max $\rm PF_v$ \\
\hline
 -1& $(601.4,98.7)$  & (1620, -473.6) & (7.5,0.75)  & (4.5,0.9) \\
  0   & (275.0,-17.0) & (3.3,1.8) & (3.8,2.2)  &  (3.3,1.8) \\
1 & (27.9, -18.3) & (20.8,6.8) & (4.8,3.0)  & (4.8,3.0)\\
2 & (27.9, -19.3) & (1515.4,175.7) & (6.1,4.1)  & (6.1,4.1)\\
 \hline    
 \hline
\end{tabular}
\end{table}

Maximum values of $\rm PF$ and $ZT$ are located beyond the range of plot in Fig.~\ref{fig10} except for max(PF) of $r=0$ which has the same location as max ($\rm PF_v$). These maximum values of $\rm PF$ and $ZT$ are located in the far right corner or the left corner of Fig.~\ref{fig10} with both large $\gamma$ and $|\xi|$. However, in these regions, SVEF is very small and even less than one. Optimal product of SVEF and TE performance yields $\rm PF_v$ and $ZT_{\rm v}$ with typical values of SVEF is around unity. We note that the larger $r$ values, the achievable TE performance increase as shown in Appendix B (Figs.~\ref{fig11} and \ref{fig12}).

\section{CONCLUSION}
 We found optimal parameters to achieve large $ZT$ and $\rm PF$ as well as their corresponding spin-valve enhancement factor (SVEF) in the half-metal ferromagnet. The large SVEF is achieved when the degree of metalicity of the HMF is large (large $E_0$). On the other hand, to obtain large PF and $ZT$, one needs to open the band gap of HMF ($E_0<\Delta$). The mismatch in these two optimized parameters indicates that spin-valve enhancement is effective only in pure HMF without band gap, in which the TE performance is rather poor. However, the achievable $ZT_{\rm v}$ and $\rm PF_v$ cannot be as large as those with the band gap. We also found that increasing the exponential factor $r$ in the relaxation time $\tau\propto (E/k_BT)^r$ also increases the TE performance. However, this does not change the fact that the resulting optimal values of $ZT_{\rm v}$ and $\rm PF_v$ cannot be higher than the highest value of $ZT$ and $PF$ in gaps. 
\begin{acknowledgments}
EHH  acknowledges financial support from the National Research Fund Luxembourg under Grants  C21/MS/15752388/NavSQM. ABC acknowledges support from Universitas Indonesia through PUTI Grant. IA acknowledges grant support from School of Electrical Engineering and Informatics, Bandung Institute of Technology's through Penelitian, Pengabdian kepada Masyarakat, dan Inovasi (PPMI) program. MSM acknowledges financial support from the e-Asia  grant 4554/IT2.IV.1.2.1/T/TU.00.08/2023.
\end{acknowledgments}

\appendix

\section{Thermoelectricity of 2D half metal ferromagnetic}
To analyze the thermoelectric coefficient of the half-metallic band structure illustrated in Fig.\ref{fig1}(b), we applied Boltzman transport theory with 
\begin{align}\label{velocity}
    v^2(E)&= \biggl( \frac{E(\textbf{k})}{m}\biggl),\\ \label{rtime}
    \tau(E)&= \tau_0 \biggl(\frac{E(\textbf{k})}{k_BT}\biggl)^r ,\\ \label{dos}
    \mathcal{D}(E)&= \frac{m}{\pi L \hbar^2}\Theta(E),
\end{align}
where $E(\textbf{k}) = \hbar^2\textbf{k}^2/2m$. Thermoelectric kernel using linearized Boltzmann transports
\begin{align}
   \mathcal{L}_i = \frac{\tau_0}{\pi L \hbar^2(k_B T)^{r}} 
   &\int E(\textbf{k})^{r+1}(E - \mu)^i\biggl(-\frac{\partial f_0}{\partial E}\biggl) dE .
\end{align}
Next, we use $\frac{\tau_0}{\pi L \hbar^2} = C,$ $E/k_B T = \epsilon$, and $\mu/k_B T = \eta$ to arrive at an analytical form

\begin{align}\label{kernelGeneral}
    \mathcal{L}_i = C (k_B T)^{i+1} \int \biggl(\frac{E(\textbf{k})}{k_BT}\biggl)^{r+1}  \frac{(\epsilon - \eta)^i \exp{(\epsilon-\eta)}}{[\exp{(\epsilon-\eta)+1}]^2} d\epsilon.
\end{align}

\subsubsection{Insulating band kernel}
 Constant relaxation time approximation ($r=0$), with dispersion energy $E_c(\textbf{k}) = E(\textbf{k}) - \Delta$ for conduction band, while $E_v(\textbf{k}) = -E(\textbf{k}) - \Delta$ for valence band, $(\tilde{\Delta} = \Delta /k_B T)$ 
 \begin{align}
     \mathcal{L}_{(c,v),i} = C (k_B T)^{i+1} \int \frac{(\pm\epsilon - \tilde{\Delta})(\epsilon - \eta)^i  \exp{(\epsilon-\eta)}}{[\exp{(\epsilon-\eta)+1}]^2} d\epsilon.
 \end{align}
Here, we introduced $\epsilon-\eta = x$,  $d\epsilon = dx $, with $i = 0,1,2$
\begin{align}\nonumber
      \mathcal{L}_{(c,v),i} &= C(k_B T)^{i+1} \int  \frac{x^i ( \pm(x +\eta) - \tilde{\Delta}) \exp{(x)}}{[\exp{(x)+1}]^2} dx\\ 
&= C(k_B T)^{i+1} \biggl( \pm\int  \frac{x^{i+1}\exp{(x)}}{[\exp{(x)+1}]^2} dx\nonumber\\ 
& +(\pm\eta-\bar{\Delta}) \int  \frac{x^{i}\exp{(x)}}{[\exp{(x)+1}]^2} dx\biggl)  ,
\end{align}
for conduction band, energy range $\Delta\leq E<\infty$, and valence band $\Delta\leq-E<\infty$. So the integral boundaries $\bar{\Delta}+\eta\leq x<\infty$ for conduction kernel, and $-\infty<x\leq -\bar{\Delta}-\eta$ for valence band kernel.

\begin{align}
      \mathcal{L}_{c,i} =& C(k_B T)^{i+1} \biggl(\int_{\Tilde{\Delta}-\eta}^{\infty}  \frac{x^{i+1}\exp{(x)}}{[\exp{(x)+1}]^2} dx\nonumber\\ 
      & +(\eta-\bar{\Delta}) \int_{\Tilde{\Delta}-\eta}^{\infty}   \frac{x^{i}\exp{(x)}}{[\exp{(x)+1}]^2} dx\biggl)  ,
\end{align}

\begin{align}
      \mathcal{L}_{v,i} =& C(k_B T)^{i+1} \biggl(-\int_{-\infty}^{-\Tilde{\Delta}-\eta}  \frac{x^{i+1}\exp{(x)}}{[\exp{(x)+1}]^2} dx\nonumber\\ 
      &
      +(-\eta-\bar{\Delta}) \int_{-\infty}^{-\Tilde{\Delta}-\eta}   \frac{x^{i}\exp{(x)}}{[\exp{(x)+1}]^2} dx\biggl)  .
\end{align}

Using the following integrals
\begin{align}
    \mathcal{F}_{i,c}(\Delta) = \int_{\Delta}^{\infty}\frac{x^i\exp{x}}{[\exp{(x)+1}]^2}dx
\end{align}
and
\begin{align}
    \mathcal{F}_{i,v}(\Delta) = -\int_{-\infty}^{-\Delta}\frac{x^i\exp{x}}{[\exp{(x)+1}]^2}dx,
\end{align}
conduction and valence band kernels can be written as
\begin{align}
    \mathcal{L}_{c,i}=C(k_B T)^{i+1}\biggl( \mathcal{F}_{i+1,c}(\Tilde{\Delta}-\eta) - (\Tilde{\Delta}-\eta)\mathcal{F}_{i,c}(\Tilde{\Delta}-\eta) \biggl)
\end{align}
and 
\begin{align}
    \mathcal{L}_{v,i}=C(k_B T)^{i+1}\biggl( \mathcal{F}_{i+1,v}(\Tilde{\Delta}+\eta) + (\Tilde{\Delta}+\eta)\mathcal{F}_{i,v}(\Tilde{\Delta}+\eta) \biggl),
\end{align}
respectively.

Integral forms on $\mathcal{F}$ can be analytically evaluated to give the following analytic forms. For conduction band:
\begin{align}
\label{f0}
 \mathcal{F}_{o,c}(x) =& \frac{1}{1 + \rm{e}^x},\\
 \label{f1}
 \mathcal{F}_{1,c}(x) =& -\frac{x\rm{e}^x}{1 + \rm{e}^x} + \rm{ln}(1+\rm{e}^x),\\
\label{f2}
 \mathcal{F}_{2,c}(x) =& -\frac{x^2\rm{e}^x}{1 + \rm{e}^x} + \frac{\pi^2}{3} + 2x \rm{ln}(1+\rm{e}^x) + 2\rm{Li}_2(-\rm{e}^x),\\
 \label{f3}
 \mathcal{F}_{3,c}(x) =&  -\frac{x^3\rm{e}^x}{1 + \rm{e}^x} + 3x^2 \rm{ln}(1+\rm{e}^x) +6x \rm{Li}_2(-\rm{e}^x) - 6\rm{Li}_3(-\rm{e}^x).
\end{align}
On the other hand, integral for the valence band, obey $
\mathcal{F}_{i,v}(x) = (-1)^{i+1}\mathcal{F}_{i,c}((-1)^{i}x)
$, where $i=0,1,2,3.$

\subsubsection{Metallic band kernel}
Similar to the insulating band, the thermoelectric kernel for the metallic band has an energy range $-E_0\leq E<\infty$, and introduced $-\bar{E_0}-\eta\leq x<\infty$ in the kernel 

\begin{align}
      \mathcal{L}_{m,i} =& C(k_B T)^{i+1} \biggl(\int_{-\bar{E_0}-\eta}^{\infty}  \frac{x^{i+1}\exp{(x)}}{[\exp{(x)+1}]^2} dx\nonumber \\
      &+(\eta+\bar{E_0}) \int_{-\bar{E_0}-\eta}^{\infty}   \frac{x^{i}\exp{(x)}}{[\exp{(x)+1}]^2} dx\biggl)\\
      =& C(k_B T)^{i+1} \biggl(\mathcal{F}_{i+1,m}(\eta+\bar{E_0}) \nonumber\\
      &+(\eta+\bar{E_0}) \mathcal{F}_{i,m}(\eta+\bar{E_0})\biggl),
\end{align}
where,
\begin{align}
    \mathcal{F}_{i,m}(\bar{E_0}) = \int_{-\bar{E_0}}^{\infty}   \frac{x^{i}\exp{(x)}}{[\exp{(x)+1}]^2} dx= \mathcal{F}_{i,c}(-E_0).
\end{align}



\section{Energy-dependent Relaxation Time}

\begin{figure*}[t]
\centering
\includegraphics[width=16cm]{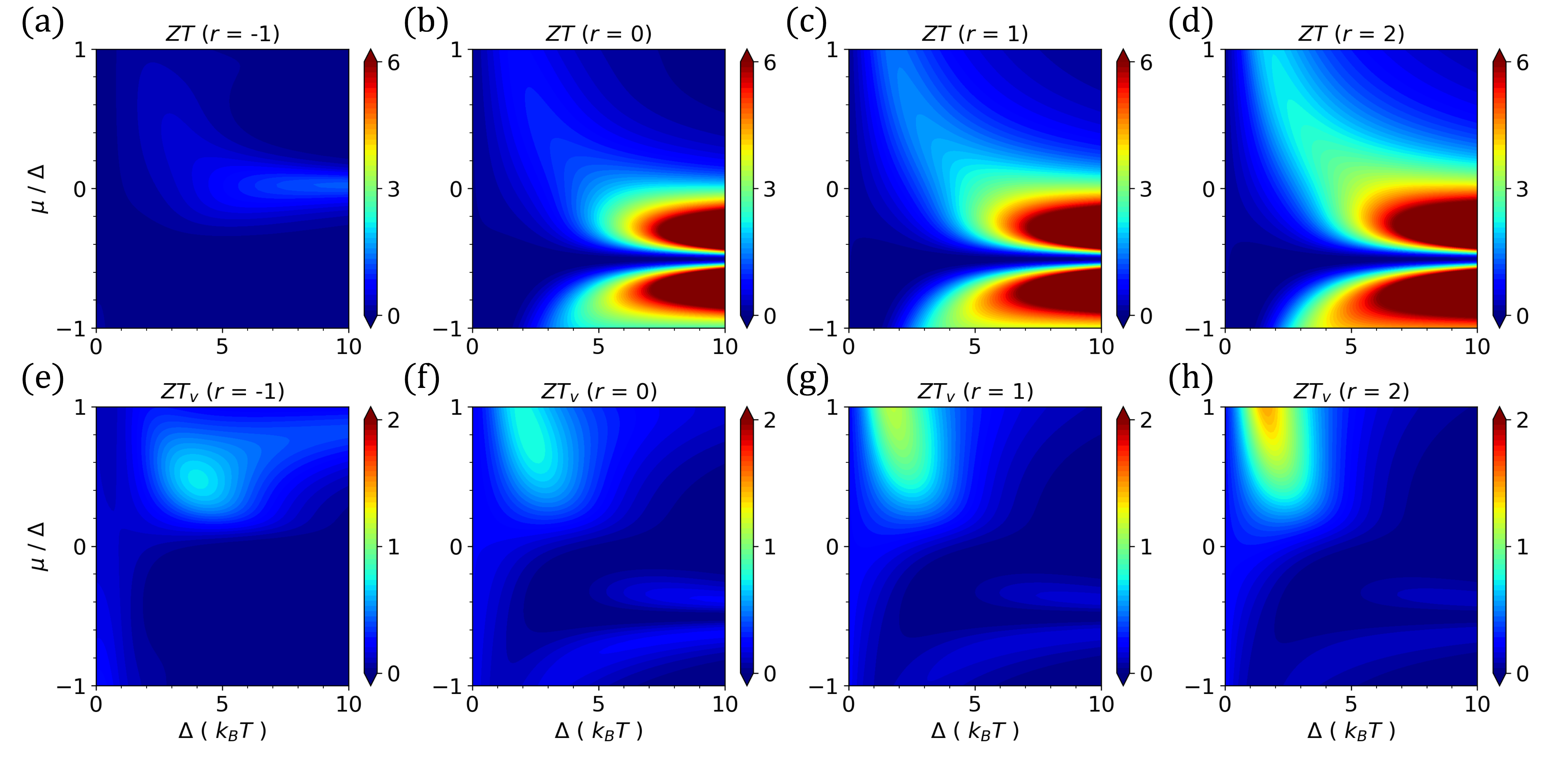}
\caption{
(a)-(d) and (e)-(h) show $ZT$ and $ZT_v$, respectively, for various $r$ from $-1$ to $2$ as function of Fermi energy $\mu$ and $\Delta$ for $E_0=0.1k_BT$
}\label{fig11}
\end{figure*}

\begin{figure*}[t]
\centering
\includegraphics[width=16cm]{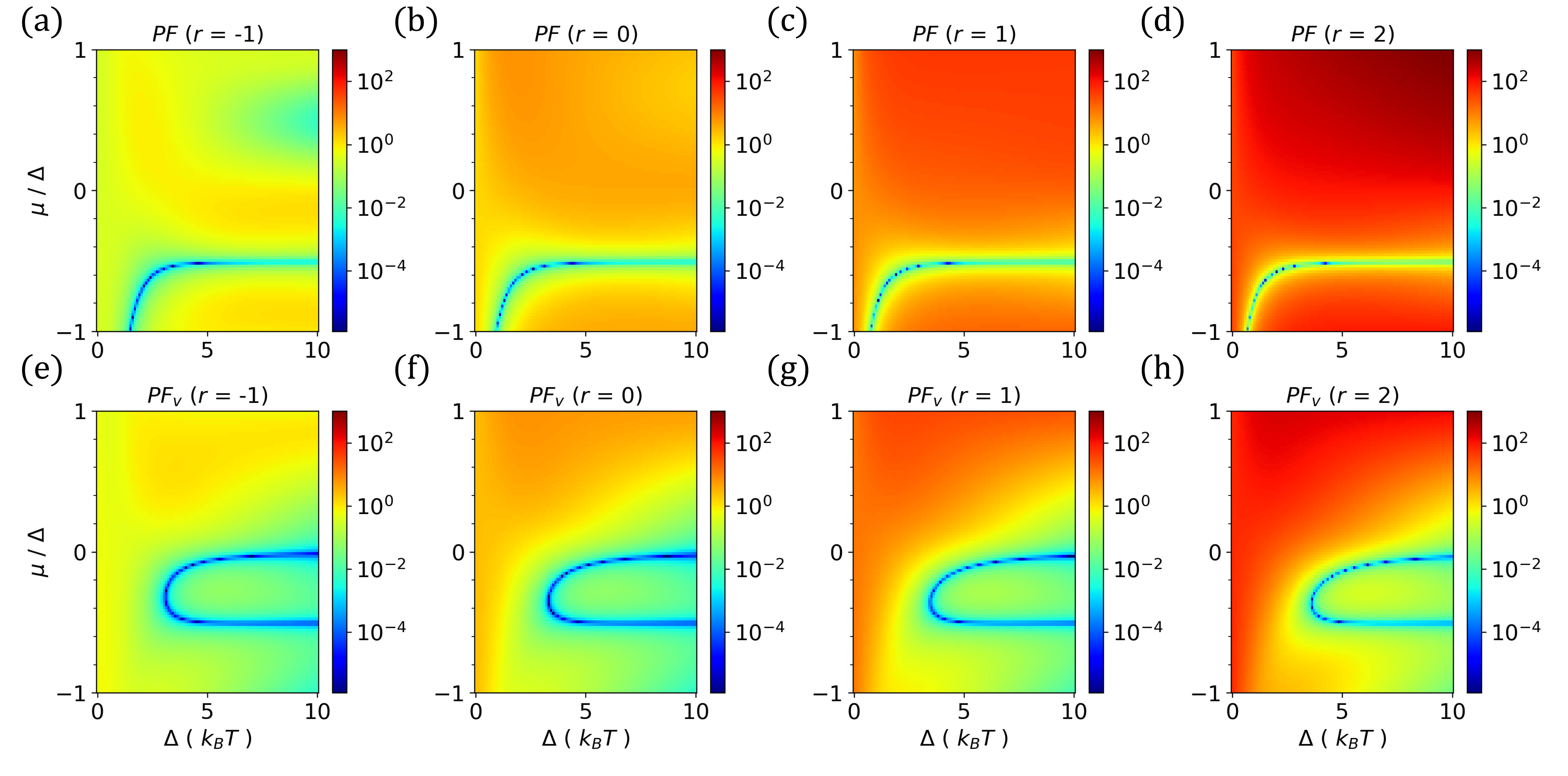}
\caption{
(a)-(d) and (e)-(h) show $\rm PF$ and ${\rm PF}_v$, respectively, for various $r$ from $-1$ to $2$ as function of Fermi energy $\mu$ and $\Delta$ for $E_0=0.1k_BT$
}\label{fig12}
\end{figure*}

Dependence relaxation time of to energy is calculated using equation \ref{rtime}. We calculate figure of merit and power factor for different $r$ as shown in figure \ref{fig11},\ref{fig12}. Thermoelectric kernel for each relaxation time are calculated using eq.\ref{kernelGeneral}. The results show that relaxation time changes influence the $ZT$ enhancement by SVEF. When relaxation time is inverse to energy, SVEF can improve the $ZT$ of HMF. This is the same as shown in figure \ref{fig10}, where the circle markers tend to be located at regions with high SVEF. Power factor in figure \ref{fig12} shows that $r$ change improves the power factor of HMF as same with  the $ZT$, at the same time SVEF is decreased.

\bibliographystyle{chem-acs}
\end{document}